\documentclass[twocolumn,showpacs,prb,floatfix]{revtex4}

\usepackage{graphicx}
\usepackage{dcolumn}
\usepackage{bm}

\usepackage{epsfig}

\def\be{\begin{equation}}
\def\ee{\end{equation}}
\def\bes{\begin{equation*}}
\def\ees{\end{equation*}}
\def\bea{\begin{eqnarray}}
\def\eea{\end{eqnarray}}
\def\beas{\begin{eqnarray*}}
\def\eeas{\end{eqnarray*}}

\begin{document}


\title{Competing Compressible and Incompressible Phases in Rotating Atomic Bose Gases at Filling Factor $\nu=2$}

\author{N. R. Cooper$^{1}$ and E. H. Rezayi$^{2}$}

\affiliation{$^1$ T.C.M. Group, Cavendish Laboratory, J. J. Thomson Avenue,
Cambridge, CB3 0HE, United Kingdom.\\ $^2$ Department of Physics, California
State University, Los Angeles, California 90032.}

\date{17 October 2006}

\begin{abstract}

We study the groundstates of weakly interacting atomic Bose gases
under conditions of rapid rotation.
We present the results of large-scale exact diagonalisation studies on
a periodic geometry (a torus) which allows studies of compressible
states with broken translational symmetry.
Focusing on filling factor $\nu=2$, we show a competition between the
triangular vortex lattice, a quantum smectic state, and the
incompressible $k=4$ Read-Rezayi state. We discuss the corrections
arising from finite size effects, and the likely behaviour for large
system sizes.  The Read-Rezayi state is stabilised by a moderate
amount of additional dipolar interactions.

\end{abstract}

\pacs{03.75.Lm, 03.75.Kk, 73.43.Cd, 73.43.Nq}

\maketitle

\section{Introduction}

Under conditions of rapid rotation, ultra-cold gases of bosonic atoms
are predicted to enter a very interesting regime of
strongly-correlated many-particle physics.\cite{wgs} At sufficiently
high vortex density, quantum fluctuations of the vortices can drive a
quantum phase transition\cite{cwg,shm1,baymmelt} from the condensed
vortex lattice phase into strongly-correlated liquid phases which are
bosonic versions of conventional and unconventional fractional quantum
Hall states.\cite{wgs,wgcw,cwg,RegnaultJ,chang:013611}
The condition for quantum melting of the vortex lattice was first
discussed in Ref.\onlinecite{cwg}, where the ``filling factor'',
$\nu$, was shown to be the relevant control parameter, and exact
diagonalisation studies were used to show evidence of a triangular
vortex lattice for $\nu \gtrsim 6$.  (For a uniform system, $\nu
\equiv N/N_{\rm v}$ where $N$ is the number of bosons and $N_{\rm v}$ the number
of vortices.)

While the limits of large\cite{cwg,shm1,baymmelt,LiebS06} and
small\cite{wgs,wgcw,cwg,RegnaultJ,chang:013611} filling factors are
well understood, the nature of the groundstates in the regime of
intermediate filling factors is less clear.  This is a difficult
regime to study theoretically, but is likely to show some of the most
interesting new physics and will be the first to be accessed as
experiments progress towards the strongly-correlated
regime\cite{expt1,expt2}. In Ref.\onlinecite{cwg} evidence was
presented for the appearance of the sequence of Read-Rezayi (RR)
states\cite{ReadR99} at $\nu=k/2$ with $k$ integer, which are
incompressible liquid states whose quasiparticle excitations obey
non-abelian exchange statistics.\cite{MooreR91} Subsequent exact
diagonalisation studies have shown that good convergence to the RR
states at $\nu\geq 3/2$ is not achieved with available system
sizes.\cite{RegnaultJ,rrc} The role of the RR state at $\nu=3/2$ was
significantly clarified in Ref.\onlinecite{rrc} where it was shown
that this state provides an excellent description of the groundstate
when a small additional non-local interaction is introduced; similar
effects have been reported for $\nu=2$ and $5/2$ in the spherical
geometry.\cite{RegnaultJ06}

In this paper we study the groundstates of rotating bosons in the
regime of intermediate filling factor using exact diagonalisation
studies in the torus geometry up to system sizes much larger than
those previously reported in this regime. We show that the
groundstates involve a competition between compressible states with
broken translational symmetry and the incompressible RR states.  Our
results identify a competing ``smectic'' phase, and point out the
important role that finite size effects can have in stabilising this
state both in numerical studies and in experimental systems.

\section{Model}

We consider a system of bosonic atoms of mass $M$ confined to a
harmonic trap with cylindrical symmetry, and with frequencies
$\omega_\parallel$ and $\omega_\perp$ in the axial and transverse
directions setting the lengthscales $a_{\parallel,\perp}\equiv
\sqrt{{\hbar}/({M\omega_{\parallel,\perp}})}$.
The atoms interact through contact interactions $V({\bm r}) = g
\,\delta^3({\bm r})$, with $g = 4\pi \hbar^2 a_s/M$ chosen to reproduce
the $s$-wave scattering length $a_s$.
We study the regime of weak interactions, when the mean interaction
energy is small compared to the trap energies, such that the particles
are restricted to the quasi-2D lowest Landau level (LLL)
regime.\cite{wgs,ButtsR99Ho01,expt1}  In this limit, the
characteristic energy scale is set by the $s$-wave Haldane
pseudopotential, $V_0 = {g}/{[(2\pi)^{3/2} a_\perp^2a_\parallel]}$.
We have investigated the nature of the bulk groundstates ({\it i.e.} of systems
containing large numbers of vortices\cite{footnote1}) using exact
diagonalisation studies on a periodic rectangular geometry (a torus)
of sides $a$ and $b$, which contains $N_{\rm v} = ab/(\pi a_\perp^2)$
vortices.\cite{cwg}
The torus lends itself naturally to the present study: it is
commensurate with crystalline order so allows the study of states with
broken translational invariance; the existence of a conserved
momentum\cite{haldanemtm} reduces the sizes of the matrices in the
diagonalisations and permits the identification of
translational-symmetry-broken states through the appearance of
quasi-degeneracies at wavevectors equal to the reciprocal lattice
vectors (RLVs) of the crystalline order.\cite{RezayiHY99}  We express
the momentum as a dimensionless wavevector ${\bm K} = (K_x,K_y)$ using
units of $2\pi\hbar/a$ and $2\pi\hbar/b$ for the $x$ and $y$
components, and report only positive $K_x,K_y$ [states at $(\pm
K_x,\pm K_y)$ are degenerate by symmetry].

\section{Results}

We find that for filling factors $\nu\geq 2$ the spectrum is sensitive
to the aspect ratio $a/b$ of the torus.\cite{footnote1} The sensitivity to
boundary conditions is a sign of a compressible groundstate, at
least on the lengthscale defined by the finite system size in the
calculation.  Here, we shall focus on filling factor $\nu=2$, which is
convenient to study from a numerical point of view.  (The qualitative
features are reproduced at larger filling factors.)

\begin{figure}
\includegraphics[width=7.5cm]{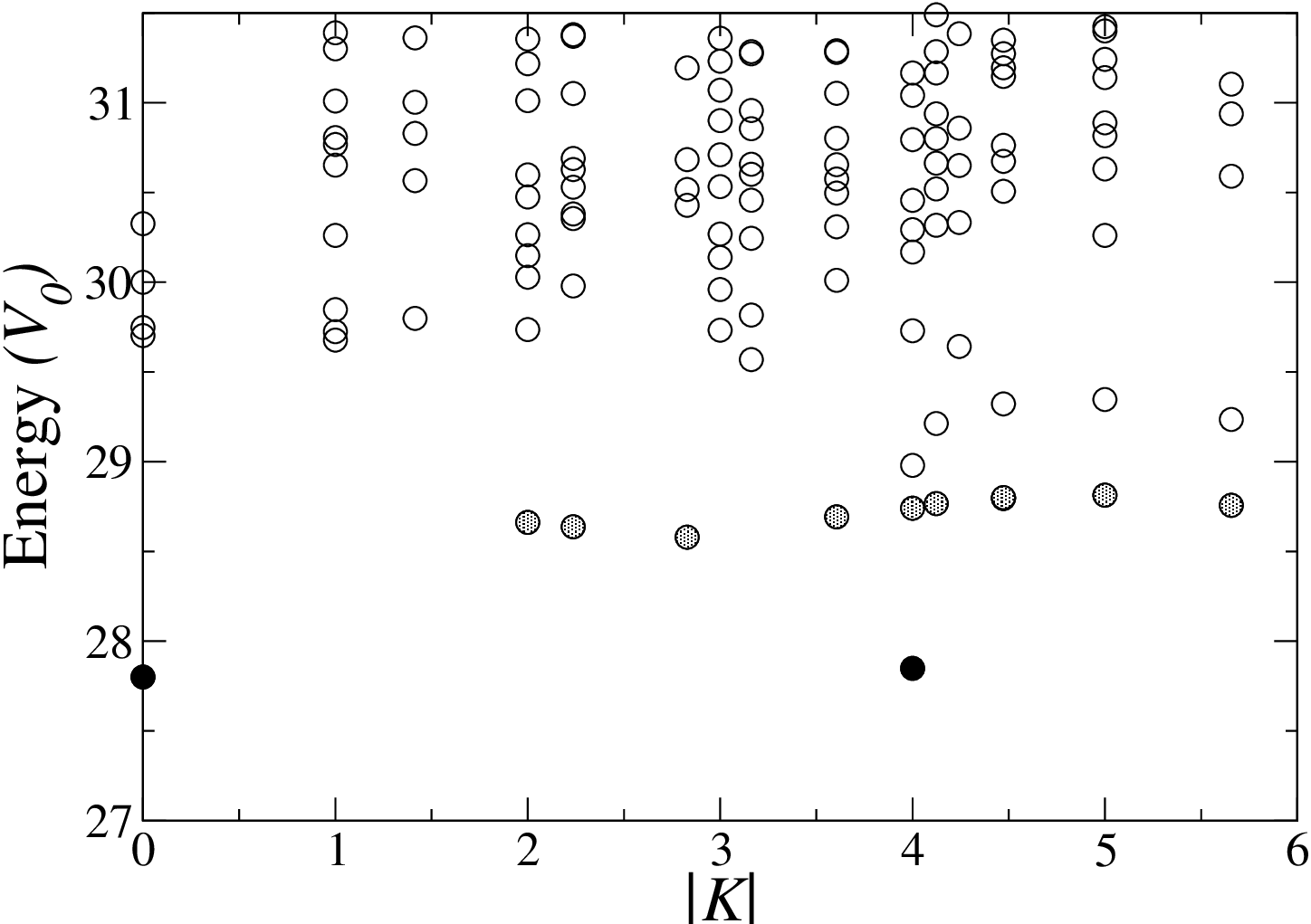}
\includegraphics[width=7.5cm]{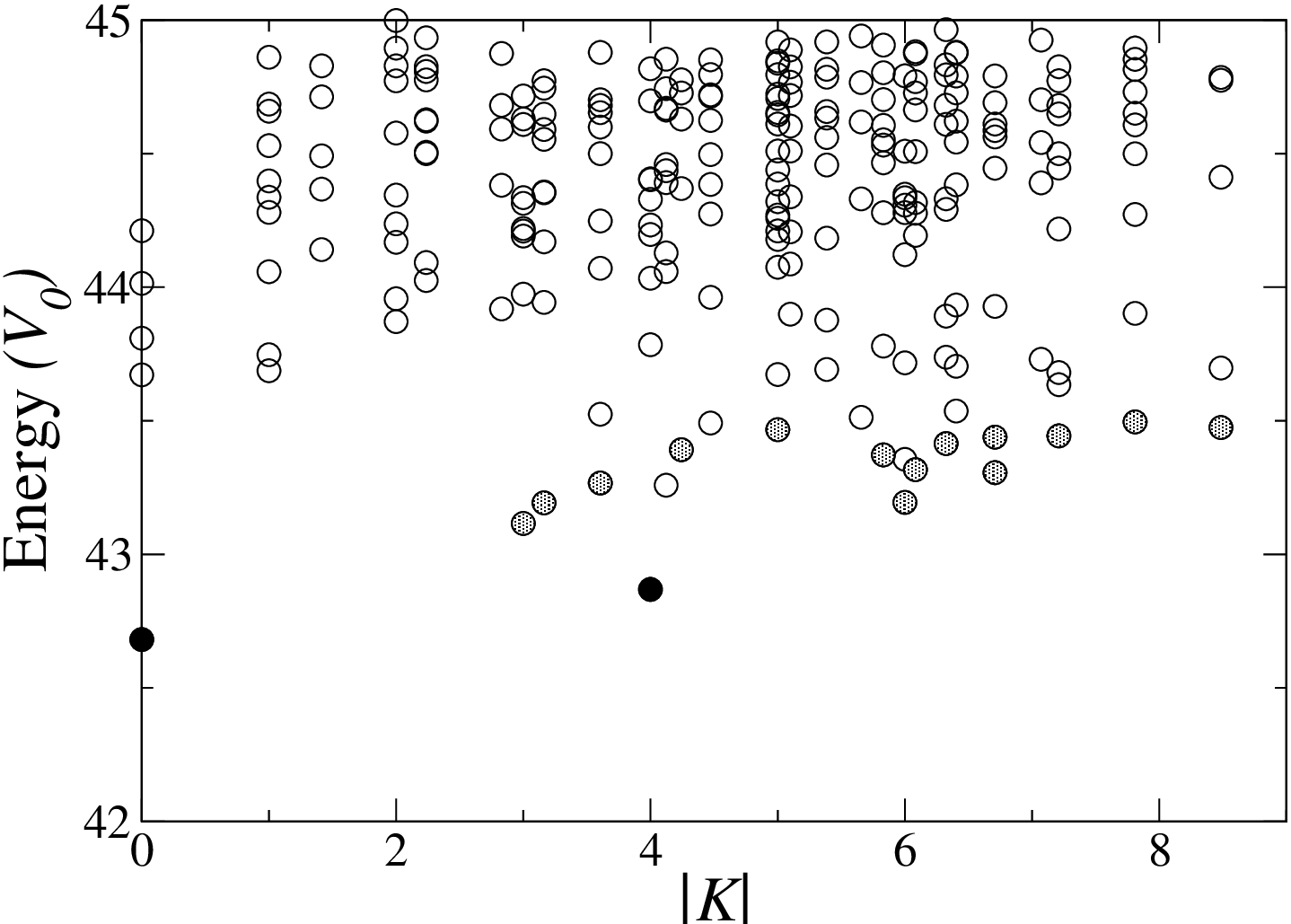}

\caption{Energy spectrum at $\nu=2$ as a function of
$\sqrt{K_x^2+K_y^2}$, with torus aspect ratio chosen to be close to
the value for which the groundstate energy is a local minimum. (a)
$N_{\rm v}=8, N=16, a/b=2$ (b) $N_{\rm v}=12, N=24, a/b=4/3$.  The
solid symbols mark the two lowest-energy states indicating the
ordering wavevector of the smectic groundstate. The shaded symbols
mark a band of excitations discussed in the text. [Note that the
components $K_x,K_y$ are expressed in units of $2\pi/a$ and
$2\pi/b$.]}

\label{fig:excitation}
\end{figure}

\subsection{Smectic Groundstate}

In Fig.~\ref{fig:excitation} we present the low-energy spectra at
$\nu=2$ for systems containing $N_{\rm v}=8$ and $12$ vortices, at
aspect ratios ($a/b=2$ and $4/3$) which are very close to the values
for which the groundstate energy has a local minimum.\cite{footnote2}
In both cases, the groundstate [at ${\bm K}=(0,0)$] is
quasi-degenerate with a single state at non-zero wavevector [at ${\bm
K}=(4,0)$]. This indicates a tendency to the formation of a {\it
smectic} groundstate\cite{ChaikinLFradkinK99} -- {\it i.e.}  a density
wave state in which translational symmetry is broken in only one
direction.  In Figs.\ref{fig:excitation}(a) and (b) the smectic
groundstates consist of four stripes parallel to the $y$-axis, with
$4$ and $6$ particles per stripe respectively.  Although the smectic
period is close to the lattice constant of a square lattice, $a^{\rm
sq} = \sqrt{\pi} a_\perp$, there is no indication of ordering in a
vortex lattice (there are no quasi-degenerate states at the requisite
RLVs).

The low-energy excitation spectra above these quasi-degenerate states
are consistent with the existence of a smectic groundstate.
For $N_{\rm v}=8$, Fig.~\ref{fig:excitation}(a), the groundstate has
four stripes of four particles each. There exist low-energy
``particle-hole'' excitations in which a particle is moved from one
stripe to another (to give occupations of $4,4,3,5$). These
excitations account for the band of states shown as shaded symbols in
Fig.~\ref{fig:excitation} (a). This band is narrow showing that
exchange interactions between the stripes are small.  For larger
systems (longer stripes) one expects the energy of this band to fall
and exchange interactions to increase, so these particle-hole
excitations will contribute increasingly to the groundstate.  It is
possible that this will lead to a phase locking of the stripes,
causing the formation of a vortex lattice: a set of states at the RLVs
of the lattice would then emerge from this band and become
quasi-degenerate with the groundstate; the remaining states would
contribute to the phonon mode of the lattice. The exact
diagonalisation studies show that, at least up to $N_{\rm v}=12$, a
vortex lattice does not form. Rather, the system shows behaviour
expected of a smectic groundstate: the analogous low-energy band
[shaded symbols in Fig.~\ref{fig:excitation}(b)] broadens to give
low-energy states at ${\bm K}=(0,3)$ and $(0,6)$ corresponding to the
competing {\it re-oriented} smectic state (with three stripes of eight
particles parallel to the $x$-axis); the remaining low-energy states
in this band form the expected gapless shear mode of this
smectic state.
Thus, the exact diagonalisation results up to $N_{\rm v}=12$ are consistent
with a smectic groundstate at $\nu=2$.\cite{footnote3}

\subsection{Triangular Vortex Lattice}

\begin{figure}
\includegraphics[width=7.5cm]{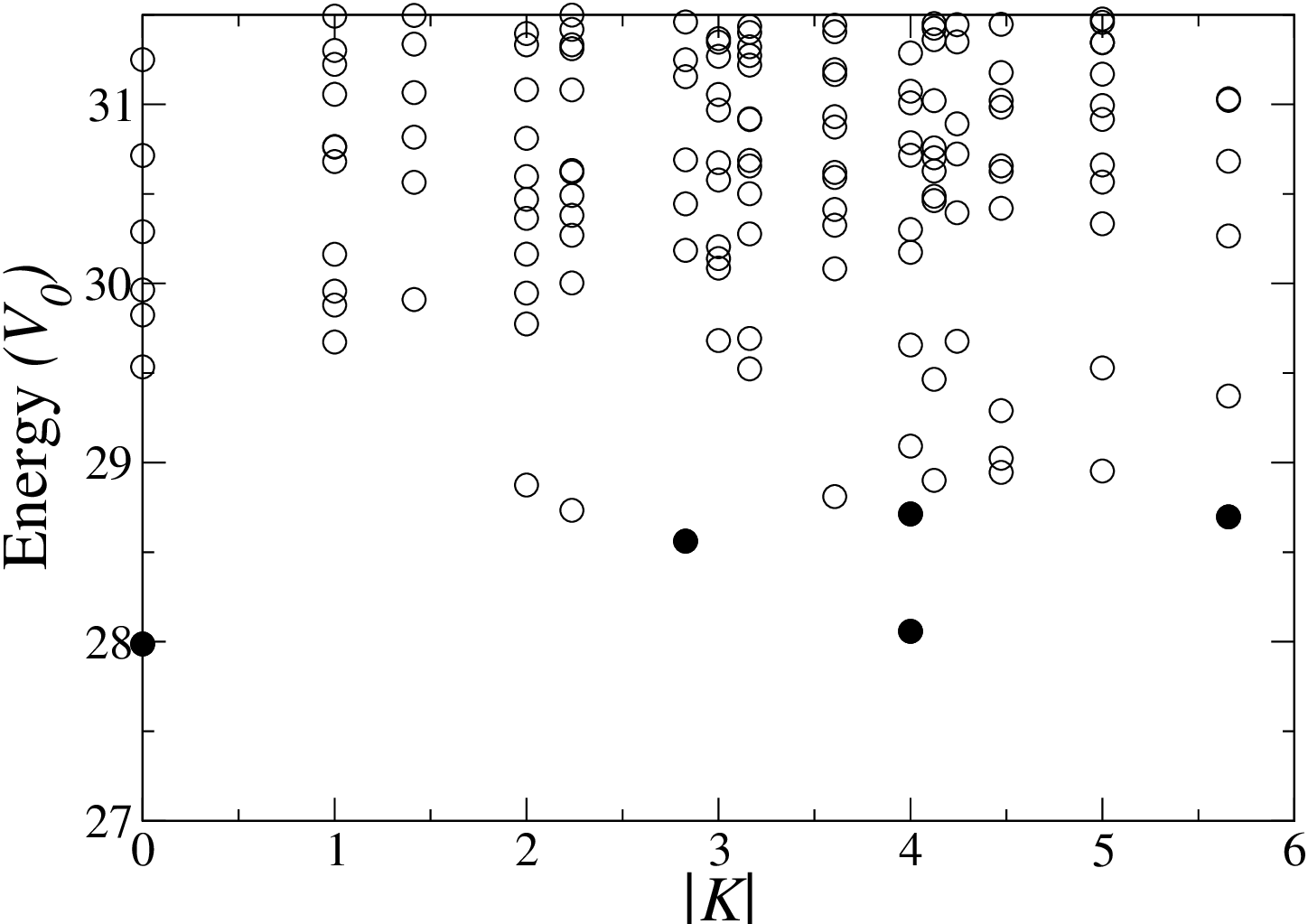}
\includegraphics[width=7.5cm]{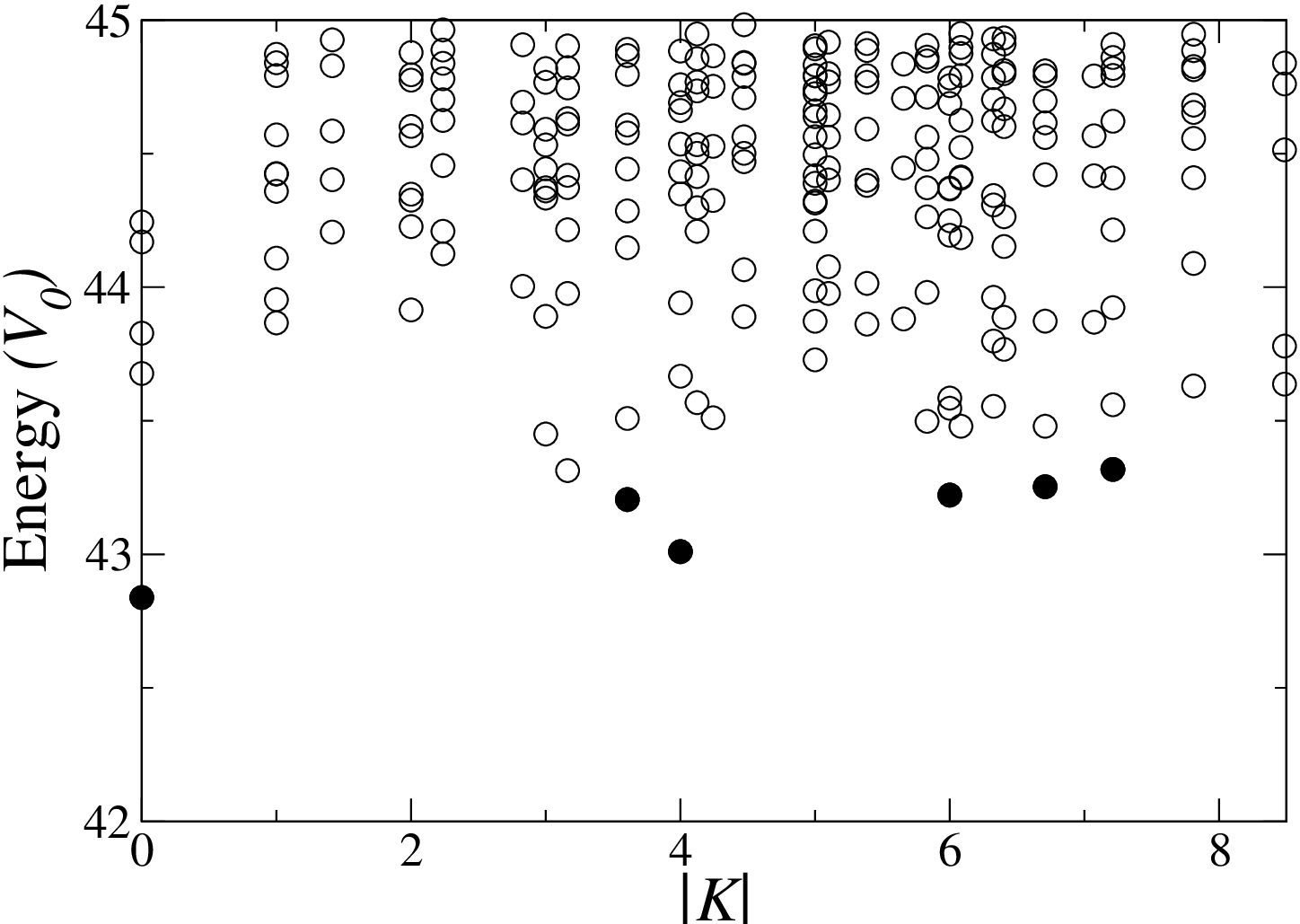}
\caption{Energy spectrum at $\nu=2$ as a function of
$\sqrt{K_x^2+K_y^2}$, with torus aspect ratio chosen to be consistent
with a triangular vortex lattice. (a) $N_{\rm v}=8, N=16,
a/b=\sqrt{3}$ (b) $N_{\rm v}=12, N=24, a/b=2/\sqrt{3}$.  The solid
symbols mark the wavevectors of a triangular vortex lattice. These
states are not well separated from higher energy states, showing a
weak tendency to vortex lattice ordering.  [Note that the components
$K_x,K_y$ are expressed in units of $2\pi/a$ and $2\pi/b$.]}
\label{fig:triangular}
\end{figure}

Evidence for a competing vortex lattice state at $\nu=2$ can be
obtained by choosing an aspect ratio that is commensurate with the
triangular vortex lattice (away from the local minimum in groundstate
energy). There then develop approximate quasi-degeneracies of the
groundstate with states at the RLVs of the triangular lattice, which
emerge from the band of low-energy states discussed above.
The spectra are shown in Fig.\ref{fig:triangular} for systems with $N_{\rm v}=8,12$.
In all systems we have studied the states at
the RLVs of the triangular lattice do not become well separated from
other states in the low-energy band. Moreover, since this aspect ratio
does not correspond to a local minimum of the energy, there is poor
evidence that the groundstate is a triangular vortex lattice.

\subsection{Finite-size Effects}

As we now discuss, finite size corrections may be acting
to favour the smectic state as compared to the triangular vortex
lattice.  We illustrate the relevance of finite size corrections by
making simple mean-field ans{\" a}tze for the triangular vortex
lattice and smectic states.

The triangular vortex lattice is the fully condensed state in the 2D
LLL that minimises the mean interaction energy.  The mean energy of
this state on a periodic geometry is
\be
\frac{E^{\rm tri}}{N} 
= \beta_{\rm A} \nu V_0
- \frac{\beta_{\rm A}}{N_{\rm v}} V_0
\ee
where $\beta_{\rm A} = 1.1596$,\cite{KleinerRA64} and the correction
at finite $N_{\rm v}$ arises from the lack of self-interaction of the
particles [the mean interaction energy is proportional to $N(N-1)$].

As a simple description of the smectic state, we consider the Fock
state, defined on a long cylinder with circumference $L_y$, in which
$N_{\rm s}$ bosons occupy each of a set of Landau-gauge orbitals
(``stripes'') that are extended around the circumference of the
cylinder and that are equally spaced along the length of the cylinder
by $\Delta x$.  (This state is similar to the Tao-Thouless
state.\cite{taothouless}) The smectic state describes a fragmented
condensate, with $N_{\rm s}$ particles in each stripe and with no
fixed phase relationship between the stripes. (In the LLL, the
establishment of a fixed phase relationship between the stripes would
establish fixed positions of the vortices, {\it i.e.}  form a vortex
lattice.)
For a large circumference, $L_y\to \infty$, the mean interaction
energy is minimised, at fixed filling factor $\nu = {N_{\rm s}\pi
a_\perp^2}/({L_y \Delta x})$, for a period $\Delta x
= 1.73 a_\perp$. This is close to the period of the smectic state
appearing in numerics, of approximately $a^{\rm sq} = 1.77 a_\perp$. At
this period the energy per particle of the smectic state is
\be
\frac{E^{\rm smectic}}{N} 
= \nu \beta_{\rm s} V_0
 - \frac{(\Delta x/a^{\rm sq})}{N_{\rm v,s}} V_0 
\label{eq:smecticenergy}
\ee 
where $\beta_{\rm s} \equiv 1.17184$, and $N_{\rm v,s} \equiv N_{\rm s}/\nu$ is
the number of vortices associated with each stripe. The finite-size
correction arises from the lack of self-interaction between the $N_{\rm s}$
particles in each stripe.

Since $\beta_{\rm s} > \beta_{\rm A}$, the smectic state has higher
energy than the triangular lattice in the thermodynamic limit ($N_{\rm
v},N_{\rm v,s}\to \infty$). This is a consequence of the general
result that repulsive interactions disfavour fragmentation of a
condensate owing to the loss of exchange
interactions.\cite{nozieresfragment} It is remarkable, however, just
how small the energy difference between these states is [$(\beta_{\rm
s} - \beta_{\rm A})/\beta_{\rm A} \sim 1\%$].  For a system on a long
cylinder with finite circumference ($N_{\rm v}\to \infty$ with $N_{\rm
v,s}$ fixed) the smectic state has lower energy than the triangular
vortex lattice for
$N_{\rm v,s} \lesssim 80/\nu$.
The stabilisation of the smectic state can be viewed as a form of Mott
transition: by fixing a definite particle number on each stripe, there
is a reduction in Hartree energy at the expense of an increase in
exchange energy.  At $\nu=2$ these finite size effects cause the trial
smectic state to have lower energy than the triangular vortex lattice
up to systems of $40$ vortices per stripe.  This is far beyond the
system sizes that can be studied in exact diagonalisations. It is also
far beyond the maximum number of vortices crossing the clouds in
existing experiments\cite{expt1,expt2} (on the order of ten),
showing that finite size effects can play a significant role in
determining the groundstates at intermediate filling factors in
experimental geometries.

The simple mean-field analysis presented here is not intended to
provide quantitative corrections to the exact diagonalisation
results. This analysis neglects quantum fluctuations which reduce the
mean energies of the smectic and vortex lattice states even in the
thermodynamic limit.
Nevertheless, there are indications that finite size corrections do
play an important role in the energetics of the smectic state in the
numerics: the smectic state has a lower energy the shorter is the
length of the stripes.
For example, for $N_{\rm v}=12, a/b=4/3$
[Fig.~\ref{fig:excitation}(b)] the two high symmetry configurations
correspond to configurations with either four short stripes with
$N_{\rm v,s}=3$ parallel to the $x$-axis, or three long stripes with
$N_{\rm v,s}=4$ parallel to the $y$-axis; they have energies [obtained
from the respective ordering wavevectors, ${\bm K}=(4,0)$ and $(0,3)$]
of $E/N= 1.786 V_0$ and $E/N= 1.796 V_0$. Similarly, from results on
$N_{\rm v}=12, a/b=3$ one can obtain an estimate of the energy for
short stripes with $N_{\rm v,s} =2$ ($E/N= 1.739 V_0$). There is a
monotonic increase in the energy of the smectic state as $N_{\rm v,s}$
increases.
Owing to the small number of system sizes available, it is difficult
to extrapolate to the thermodynamic limit for either the smectic or
vortex lattice state.  However, the sizes of the energy changes of the
smectic are comparable to the separation in energy to the triangular
vortex lattice, for which estimates of the energy per particle can be
as low as $E/N=1.180V_0$.\cite{footnote4}
Thus, while the smectic state is the groundstate in systems of up to
$N_{\rm v}=12$, it is possible that finite size effects are playing a
role in stabilising the smectic state.

In the studies of Ref.\onlinecite{cwg} in which a transition to a
vortex lattice was seen in systems of $N_{\rm v}=6,8$ vortices, the
aspect ratio used was close to the minimum of the groundstate
energy. In view of the above analysis, we can now say that in those
studies\cite{cwg} the estimate of the critical filling factor $\nu
\simeq 6$ arose from a competition between the triangular vortex
lattice and the smectic state.  The finite-size corrections to the
energy of the state competing with the triangular vortex lattice were
not recognised in Ref.\onlinecite{cwg}.  From the present analysis, we
conclude that the analysis of Ref.\onlinecite{cwg} underestimated the
stability of the vortex lattice in the thermodynamic limit. The
available numerical results are consistent with the vortex lattice
being more stable than the smectic state down to as small a filling
factor as $\nu=2$.

\subsection{Read-Rezayi State}

\begin{figure}
\includegraphics[width=7.5cm]{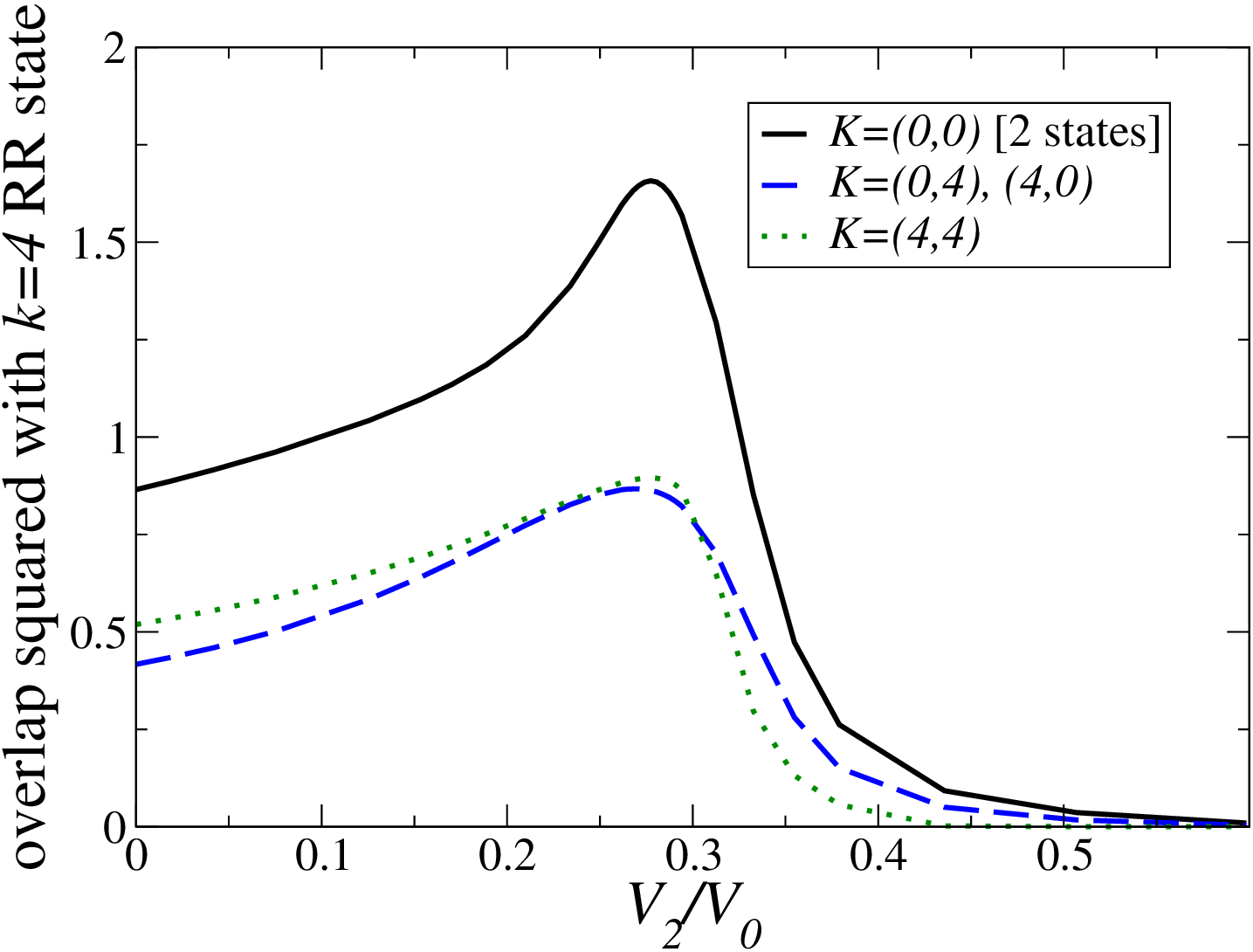}
\includegraphics[width=7.5cm]{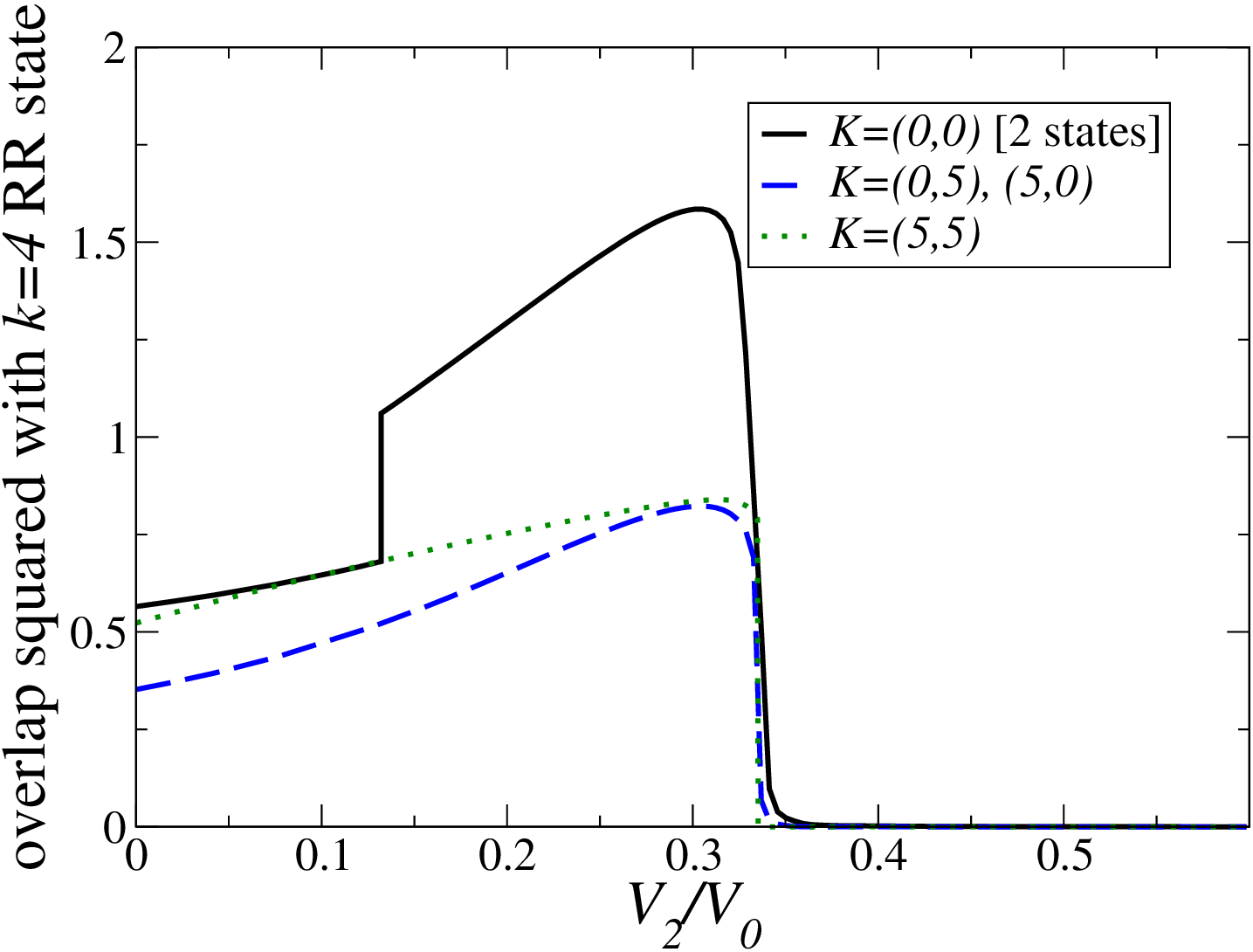}
\caption{(Color online) Variation of the overlap (squared) of the groundstate for
contact and dipolar interactions with the $k=4$ RR state, for (a)
$N_{\rm v}=8, N=16, a/b=1$ and (b) $N_{\rm v}=10, N=20, a/b=1$, for
$a_\perp/a_\parallel\ll 1$. We parameterise the relative size of
dipolar and contact interactions by the ratio of the first two Haldane
pseudopotentials $V_2/V_0$, as defined in
Refs.\onlinecite{crs,rrc}. The $k=4$ RR state has five degenerate
states on the torus located at the zone centre and
corners;\cite{ReadR99,cwg} for ${\bm K}=(0,0)$ the RR groundstate is a
doublet and we plot the overlap of the lowest two states within this
subspace, so the maximum overlap squared is $2$. [In (b), the
discontinuity of the overlap for ${\bm K}=(0,0)$ close to
$V_2/V_0=0.13$ is due to a level crossing with a state of different
point-group symmetry.]}
\label{fig:k4overlapdip}
\end{figure}

An interesting phase that competes with the smectic and the triangular
vortex lattice is the $k=4$ Read-Rezayi state.\cite{ReadR99} As
discussed above, for contact interactions the groundstate at $\nu=2$
appears to be compressible on the scale of the systems studied.
However, as for $\nu=3/2$,\cite{rrc} by introducing interactions with
non-zero range, we find that the spectrum becomes characteristic of
the incompressible RR state (showing clear quasi-degeneracies at the
expected wavevectors, and a gap to other excitations). The case of
dipolar interactions\cite{crs} is of particular experimental
relevance, as these can be important for atomic species with large
magnetic moments.\cite{GriesmaierWHSP05} 
Fig.~\ref{fig:k4overlapdip} shows the overlap of the exact
groundstate with the RR state as a function of the additional dipolar
interaction.
In a narrow range around $V_2/V_0=0.3$ the groundstate is well
described by the $k=4$ RR state.
For $V_2/V_0 \gtrsim 0.4$ the groundstate is well described by
broken-symmetry states with stripe and ``bubble crystal''
ordering.\cite{crs}  These states have periods that are larger than
$a^{\rm sq}$, and are much more stable to quantum
fluctuations\cite{crs} than are the compressible states discussed
above.
An enhancement of the overlap with the $k=4$ RR state in related
calculations on the spherical geometry for a model short-range
interaction has also been reported.\cite{RegnaultJ06}

\section{Summary}

We have shown that exact diagonalisation studies of rotating Bose
gases with contact interactions indicate that the groundstate at
$\nu=2$ is a quantum smectic state.  We argued that finite size
corrections may lead to an enhanced stability of the smectic state in
the system sizes amenable to exact diagonalisations, and the
triangular vortex lattice may have lower energy in the thermodynamic
limit. In the regime of intermediate filling factors, $\nu\geq 2$, the
correlation length of the groundstate for contact interactions remains
larger than the maximum system sizes that are available in exact
diagonalisations.  Comparisons with the RR states at $\nu=k/2$ quickly
become severely limited, as these states involve clustering of groups
of $k$ particles (at $\nu=2$ exact diagonalisations currently allow
studies of at most six clusters).  The torus is likely to favour
broken translational-symmetry states of short period, so it is
possible that the true groundstates at intermediate filling factors
are quantum states of more complex symmetry -- including quantum
nematic phases\cite{ChaikinLFradkinK99} or incompressible liquids with
small gaps.
Clearly this is a regime in which there are strong quantum
fluctuations of vortices, and where novel strongly correlated
groundstates may form.  It will be very interesting to explore this
regime experimentally.

The authors are grateful to N. Read for many helpful discussions.
This work was partially supported by the UK EPSRC Grant
No. GR/R99027/01 (N.R.C.)  and by the US DOE under contract
DE-FG03-02ER-45981 (E.H.R.).


\begin{thebibliography}{10}
\bibitem{foonote1} For a system containing a large number of vortices,
$N_{\rm v}\gg 1$, the rotation frequency $\Omega$ is close to
$\omega_\perp$. Using a Thomas-Fermi approximation for a harmonically
trapped gas with a large number of vortices, it is straightforward to
show that $\frac{\omega_\perp^2}{\Omega^2} = 1+
\frac{2\mu}{\hbar\Omega}\frac{1}{N_{\rm v}}$, so $\Omega \to
\omega_\perp$ as $N_{\rm v}\to \infty$.
\bibitem{footnote1} As discussed in Ref.\onlinecite{rrc} the state at
$\nu=3/2$ also shows sensitivity to boundary conditions, though the
features we describe in the present paper are much less evident.
\bibitem{footnote2} Similar results are found at $N_{\rm v}=6,9$ and
 $10$ and are consistent with our discussion.
\bibitem{footnote3} The remaining low-energy states in Fig.~\ref{fig:excitation}(a,b) that
are not shaded can be understood as two particle-hole excitations and
shear modes of the dominant smectic ordering.
\bibitem{footnote4} 
This estimate was obtained by averaging the energy over the RLVs of
the triangular lattice for $N_{\rm v}=12, a/b=2/\sqrt{3}$.
\end{thebibliography}

\end{document}